\documentclass[11pt]{article}

\usepackage{fullpage}
\usepackage{latexsym}

\def\01{\{0,1\}}

\newcommand{\eps}{\varepsilon} 
\newcommand{\ket}[1]{|#1\rangle}

\newtheorem{definition}{Definition}
\newtheorem{theorem}{Theorem}

\newtheorem{fact}{Fact}

\begin{document}

\title{Quantum Symmetrically-Private Information Retrieval}
\author{Iordanis Kerenidis\thanks{Supported by DARPA under agreement number F 30602-10-2-0524. Part of this work was done while visiting CWI.}\\
UC Berkeley\\
{\tt jkeren@cs.berkeley.edu}
\and Ronald de Wolf\thanks{Most of this work was done while a postdoc at UC Berkeley,
supported by Talent grant S 62--565 from the Netherlands Organization
for Scientific Research (NWO). Also (partially) funded by projects QAIP
        (IST--1999--11234) and RESQ (IST-2001-37559)
        of the IST-FET programme of the EC.}\\
CWI Amsterdam\\
{\tt rdewolf@cwi.nl}}
\date{}
\maketitle

\begin{abstract}
Private information retrieval systems (PIRs) allow a user to extract 
an item from a database that is replicated over $k\geq 1$ servers, 
while satisfying various privacy constraints.
We exhibit quantum $k$-server symmetrically-private information retrieval systems 
(QSPIRs) that use sublinear communication, do not use shared randomness among 
the servers, and preserve privacy against honest users and dishonest servers.
Classically, SPIRs without shared randomness do not exist at all.\\[2mm]
\end{abstract}

\section{Introduction}

\subsection{Setting}

The {\em Private Information Retrieval\/} problem
was introduced by Chor et~al.~\cite{cgks:pir}.
A user wants to learn a bit $x_i$ from an $n$-bit database 
$x=x_1\ldots x_n$, for some $i\in[n]$ of his choice.
The privacy of the user requires that the database server learns 
nothing about $i$, in the information-theoretic sense,
and general efficiency considerations require the communication 
between the user and the database to be limited.
Clearly, PIR can be realized by making the server send the whole 
database to the user. This takes $n$ bits of communication 
and can be shown to be optimal.
Better protocols exist if the database is {\em replicated\/}
among some $k\geq 2$ different servers, who cannot 
communicate~\cite{cgks:pir,ambainis:pir,bikr:improvedpir}.
Here we require that {\em individual\/} servers learn nothing about $i$.
For $k=2$, the best known scheme uses $O(n^{1/3})$ bits
of communication~\cite{cgks:pir}, and asymptotically the best known
$k$-server uses $n^{O(\log\log(k)/k\log(k))}$ bits~\cite{bikr:improvedpir}.
For $k\geq 2$, no good lower bounds on the required 
communication are known for this setting.

In a recent paper, we showed how to obtain {\em quantum\/} PIR
systems (QPIR, where the parties are quantum computers and the
communication consists of qubits) that use slightly less communication
than the best known classical schemes~\cite{kerenidis&wolf:qldc}.
In Table~\ref{tablepir} we list the best known bounds on 
the communication complexity for small numbers of servers, 
in the classical as well as quantum case.

\begin{table}[htb]
\centering
\begin{tabular}{|c|c|c|}
\hline
Servers & PIR complexity & QPIR complexity\\ \hline\hline
$k=1$ & $\Theta(n)$      & $\Theta(n)$ \\ \hline
$k=2$ & $O(n^{1/3})$     & $O(n^{3/10})$ \\ \hline
$k=3$ & $O(n^{1/5.25})$     & $O(n^{1/7})$ \\ \hline
$k=4$ & $O(n^{1/7.87})$     & $O(n^{1/11})$ \\ \hline
\end{tabular}
\caption{Best known bounds on the communication complexity of
classical and quantum PIR}\label{tablepir}
\end{table}

In its standard form, PIR just protects the privacy of the {\em user\/}: 
the individual servers learn nothing about $i$.
But now suppose we also want to protect the privacy of the {\em data}.
That is, we don't want the user to learn anything about $x$ beyond 
the $x_i$ that he asks for. For example, because the user should pay 
a fee for every $x_i$ that he learns (pay-per-view), or because the 
database contains very sensitive information.
This setting of {\em Symmetrically-Private\/} Information Retrieval 
(SPIR) was introduced by Gertner et~al.~\cite{gikm:pir}.
They showed that SPIR is {\em impossible\/} even if the user is honest
(i.e., follows the protocol) and the servers can individually flip 
coins~\cite[Appendix~A]{gikm:pir}. This no-go result holds no matter 
how many servers and how many bits and rounds of communication we allow.
Therefore they extended the PIR model by allowing the servers to {\em share\/}
a random string that is hidden from the user, and showed how to turn 
any PIR scheme into a SPIR scheme with shared randomness among the servers, 
at a small extra communication cost. The resulting schemes are 
information-theoretically secure even against {\em dishonest} users,
and use a number of random bits that is of the same order as the communication.

The necessity of shared randomness for classical SPIR schemes is 
a significant drawback, since information-theoretic security requires 
{\em new\/} shared randomness for each application of the scheme.
This either requires a lot of extra communication between the servers
(if new shared randomness is generated for each new application) 
or much memory on the parts of the servers (if randomness is generated
once for many applications, each server needs to store this).

\subsection{Results}

In this paper, we study the existence and efficiency of SPIR schemes 
in the {\em quantum\/} world, where user and servers have quantum 
computers and can communicate qubits. Here user privacy means that 
the states of individual servers should all be independent of $i$,
and data privacy means that the concatenation of the various states of 
the user is independent of the bits $x_j$ for all $j\neq i$.
We can distinguish between honest-user and dishonest-user data privacy.
In the first case, data privacy holds if the user is honest 
(follows the protocol). In the second case, data privacy should 
hold even if the user deviates from the protocol in any way.

Our main result is that
honest-user quantum SPIR schemes exist even in the case where the
servers do not share any randomness. As mentioned above, 
such honest-user SPIRs without shared randomness are impossible 
in the classical world.
This gives another example of a cryptographic task that can be
performed with information-theoretic security
in the quantum world but that is impossible classically 
(key distribution~\cite{bb84} is the main example of this).
The communication complexity of our $k$-server QSPIR schemes is
of the same order as that of the best known classical
$k$-server PIR schemes.  At first sight, one might think this 
trivial: just take a classical scheme, ensure data privacy 
using shared randomness among the servers, and then
get rid of the shared randomness by letting the user
entangle the messages to the servers. 
However, this would violate data privacy, as the user 
would now have ``access'' to the servers' shared randomness.
In actuality we do something quite different, making use of the fact 
that the servers can add phases that multiply out to an overall phase.
This phase allows the user to extract $x_i$, but nothing else. 
For $k=2$ we also give an alternative, less efficient scheme 
based on the properties of Bell states. 

The notion of an honest user is somewhat delicate, because
clearly users cannot be trusted to follow the protocol in all cases.
Still, there are scenarios where the assumption of a honest user is
not unreasonable, for example in pay-per-view systems where
the user accesses the system via some box (attached to his TV) 
that is sealed or otherwise protected from tampering. 
In this case the user cannot deviate from the protocol, 
but he can still be curious, trying to observe what goes on inside
of his box to try to extract more information about the database.
Our honest-user QSPIRs are perfectly secure against such users.

It would be nice to have SPIR schemes that are secure 
even against {\em dishonest\/} users.
However, we exhibit a large class of PIR schemes (quantum as well as classical) that 
can all be cheated by a dishonest quantum user. Our honest-user QSPIRs fall 
in this class and hence are not secure against dishonest users.
Fortunately, if we are willing to allow shared randomness between
the servers then the best classical SPIRs can easily be made
secure against even dishonest quantum users: if the servers measure the
communication in the computational basis, the scheme is equivalent
to the classical scheme, even if the user is quantum.

\bigskip

\noindent{\bf Remarks:}

(1) Often the PIR setting is generalized to {\em $t$-secure} PIR,
where no colluding set of $t$ servers together have any information about $i$.
We focus on the $t=1$ case here in order to simplify the presentation.

(2) Very efficient PIR and SPIR schemes are possible under 
{\em computational\/} assumptions, even for $k=1$ servers
(see e.g.~the references at~\cite{helger}).
In this paper we focus on information-theoretic security.

\section{Definitions}

We assume familiarity with the quantum model~\cite{nielsen&chuang:qc}.
The various variants of PIR are defined below.

\begin{definition}
A one-round, $k$-server private information retrieval
(PIR) scheme with recovery probability $1/2+\eps$, query size $t$,
and answer size $a$, consists of a randomized algorithm (the user),
and $k$ randomized algorithms $S_1,\ldots,S_k$ (the servers), such that
\begin{enumerate}
\item
On input $i\in[n]$, the user produces $k$ $t$-bit queries
$q_1,\ldots,q_k$ and sends these to the respective servers.
The $j$th server sends back an $a$-bit string $a_j$.
The user outputs a bit $b$ depending on $i,a_1,\ldots,a_k,$ and his randomness.
\item {\bf Recovery:}
For all $x$ and $i$, the probability (over the user's and servers' randomness)
that $b=x_i$ is at least $1/2+\eps$.
\item {\bf User privacy:}
For all $j$, the distribution of $q_j$ (over the user's randomness)
is independent of $i$.
\end{enumerate}
The {\em communication complexity\/} of the scheme is the
total length of the communication between the user and the servers, 
i.e.~$k(t+a)$ bits.
\end{definition}

All best known PIR schemes satisfy the above definitions with $\eps=1/2$ 
(i.e., no error probability), and we will hereafter take $\eps=1/2$ 
unless mentioned otherwise. It is open whether multiple-round
schemes can be better than the one-round variety we defined here.
For PIR schemes, we can assume without loss of generality that 
the servers are deterministic. However, we need randomized servers
for the symmetrically-private variety:

\begin{definition}
A {\em symmetrically-private\/} information retrieval (SPIR) scheme 
is a PIR scheme with the additional property of {\bf data privacy:}
the user's ``view'' (i.e.~the concatenation of his various states
during the protocol) does not depend on $x_j$, for all $j\neq i$.
We distinguish between {\em private-randomness\/} and {\em shared-randomness\/}
SPIR schemes, depending on whether the servers individually
flip coins or have a shared random coin (hidden from the user). 
We also distinguish between {\em honest-user\/} and 
{\em dishonest-user\/} SPIR, depending on whether data privacy 
should hold even when the user deviates from the protocol.
\end{definition}

\begin{definition}
We define quantum versions QPIR and QSPIR of PIR and SPIR, respectively, 
in the obvious way: the user and the servers are quantum computers,
and the communication uses quantum bits; 
{\em user privacy\/} means that the density matrix of each server 
is independent of $i$ at all points in the protocol; 
{\em data privacy\/} means that the concatenation of the 
density matrices that the user has at the various points of the protocol,
is independent of $x_j$, for all $j\neq i$.
For QSPIR, we still have the distinctions of 
private/public-randomness and honest/dishonest-user.
\end{definition}

As mentioned in the introduction, Gertner et~al.~\cite[Appendix~A]{gikm:pir}
exhibited quite efficient shared-randomness SPIR schemes.
One might think that these can be turned into SPIR schemes 
with deterministic servers as follows: 
the user picks a random string, sends it to each of the servers 
(along with the queries) to establish shared randomness between them,
and then erases (or ``forgets'') his copy of the random string.
However, this erasing of the random string by the user
is ruled out by the definition, since the user's view includes 
the random string he drew.
In fact, Gertner et~al.~\cite[Appendix~A]{gikm:pir}
showed that shared randomness between the servers is necessary 
for the existence of classical SPIR (even for multi-round protocols):

\begin{fact}
For every $k\geq 1$, there is no $k$-server private-randomness SPIR scheme.
\end{fact}

Intuitively, the reason is that since the servers have no knowledge of $i$ 
(by user privacy), their individual messages need to be independent of 
{\em all\/} bits of $x$, including $x_i$, to ensure data privacy.  
But since they cannot coordinate via shared randomness, their 
{\em joint} messages will be independent of the whole $x$ as well, 
so the user cannot learn $x_i$.

Below we show that this negative result does not apply to the quantum world:
using coordination via quantum entanglement, we can get honest-user 
QSPIRs without any communication or shared randomness between 
the servers at any stage of the protocol.

\section{Honest-user quantum SPIR schemes}

\subsection{Honest-user QSPIRs from PIRs}

Our honest-user QSPIR schemes work on top of the PIR schemes
recently developed by Beimel et al.~\cite{bikr:improvedpir}.
These, as well as all others known, work as follows:
the user picks a random string $r$, and depending on $i$ and $r$,
picks $k$ queries $q_1,\ldots,q_k\in\01^t$.
He sends these to the respective servers, who respond
with answers $a_1,\ldots,a_k\in\01^a$.
The user then outputs 
$$
\sum_{j=1}^k a_j\cdot b_j=x_i,
$$
where $b_1,\ldots,b_k\in\01^a$ are determined by $i$ and $r$,
and everything is modulo 2.

We will now describe the quantum SPIR scheme.
As before, the user picks $r,q_1,\ldots,q_k$.
In addition, he picks $k$ random strings $r_1,\ldots,r_k\in\01^a$.
He defines $r'_j=r_j+b_j$ and sets up the following $(k+1)$-register state
$$
\frac{1}{\sqrt{2}}\ket{0}\ket{q_1,r_1}\cdots\ket{q_k,r_k}+
\frac{1}{\sqrt{2}}\ket{1}\ket{q_1,r'_1}\cdots\ket{q_k,r'_k}.
$$
The user keeps the first 1-qubit register to himself,
and sends the other $k$ registers to the respective servers.
The $j$th server sees a random mixture of $\ket{q_j,r_j}$
and $\ket{q_j,r'_j}$.  Since $q_j$ gives no information about $i$
(by the user privacy of the classical PIR scheme) and each of 
$r_j$ and $r'_j$ is individually random, 
the server learns nothing about $i$.
The $j$th server performs the following unitary mapping
$$
\ket{q_j,r}\rightarrow(-1)^{a_j\cdot r}\ket{q_j,r},
$$
which he can do because $a_j$ only depends on $q_j$ and $x$.
The servers then send everything back to the user; the overall 
communication is $2k(t+a)$ qubits, double that of the original scheme.
The user now has the state
$$
\frac{1}{\sqrt{2}}\ket{0}(-1)^{a_1\cdot r_1}\ket{q_1,r_1}\cdots(-1)^{a_k\cdot r_k}\ket{q_k,r_k}+
\frac{1}{\sqrt{2}}\ket{1}(-1)^{a_1\cdot r'_1}\ket{q_1,r'_1}\cdots(-1)^{a_k\cdot r'_k}\ket{q_k,r'_k}.
$$
Up to an insignificant global phase $(-1)^{\sum_j a_j\cdot r_j}$, 
this is equal to 
$$
\begin{array}{ll}
\displaystyle \frac{1}{\sqrt{2}}\ket{0}\ket{q_1,r_1}\cdots\ket{q_k,r_k}+
\frac{1}{\sqrt{2}}\ket{1}(-1)^{\sum_{j=1}^k a_j\cdot b_j}\ket{q_1,r'_1}\cdots\ket{q_k,r'_k} & = \\[2mm]
\displaystyle \frac{1}{\sqrt{2}}\ket{0}\ket{q_1,r_1}\cdots\ket{q_k,r_k}+
\frac{1}{\sqrt{2}}\ket{1}(-1)^{x_i}\hspace{3.2em}\ket{q_1,r'_1}\cdots\ket{q_k,r'_k}. & 
\end{array}
$$
The user can learn $x_i$ from this by returning everything except 
the first qubit to 0, and then applying the Hadamard transform 
to the first qubit, which maps 
$\frac{1}{\sqrt{2}}\ket{0}+\frac{1}{\sqrt{2}}(-1)^{x_i}\ket{1}\rightarrow\ket{x_i}$.
On the other hand, he can learn {\em nothing else}, since the various
states of the user during the protocol never depend on any other $x_j$.
Accordingly, we have an honest-user QSPIR scheme with recovery probability 1.
Note that nowhere in the protocol do the servers have shared randomness:
they do not start with it, the random strings $r_j$, $r'_j$ are not correlated
between servers, and the servers do not end with any shared randomness
(in fact they end with nothing).

Plugging in the best known classical PIR schemes,
due to~\cite{bikr:improvedpir}, gives

\begin{theorem}
For every $k\geq 2$, there exists a honest-user QSPIR (without shared randomness)
with communication complexity $n^{O(\log\log(k)/k\log(k))}$.
\end{theorem}

Slightly better complexities can be obtained for small $k$,
as stated in the first column of Table~\ref{tablepir} in the introduction.
For $k=1$ our scheme communicates $2n$ qubits (just start from a 1-server scheme 
with query length 0, $a_1=x$ and $b_1=e_i$),
for $k=2$ it uses $O(n^{1/3})$ qubits, 
for $k=3$ it uses $O(n^{1/5.25})$ qubits etc.
Notice that we cannot use the (slightly better) $k$-server 
QPIR schemes from the second column of Table~\ref{tablepir}, 
since these reveal more than 1 bit about $x$.

\subsection{Honest-user 2-server QSPIR with Bell states}\label{secqspirk2}

The QSPIR scheme of the previous section requires communication
$O(n^{1/3})$ for the case of two servers. Here we present a different
scheme based on the Bell states. The scheme is suboptimal since it
requires linear communication, but it makes use of some interesting
properties of the Bell states and it could be easier to implement in
the lab. 
 
Our scheme works for even $n=2m$, but for odd $n$ we can just add 
a dummy bit to $x$ to make it even. It relies on three of the Bell states:
$$
\ket{B_{00}}=\frac{\ket{00}+\ket{11}}{\sqrt{2}}, \
\ket{B_{01}}=\frac{\ket{01}+\ket{10}}{\sqrt{2}}, \
\ket{B_{10}}=\frac{\ket{00}-\ket{11}}{\sqrt{2}}
$$
and the four Pauli matrices
$$
\begin{array}{llllll}
\sigma_{00} & = & \left(\begin{array}{rr}1 & 0\\ 0 & 1\end{array}\right), \
& \sigma_{01} & = & \left(\begin{array}{rr}0 & 1\\ 1 & 0\end{array}\right),\\[3mm]
\sigma_{10} & = & \left(\begin{array}{rr}1 & 0\\ 0 & -1\end{array}\right), \
& \sigma_{11} & = & \left(\begin{array}{rr}0 & -1\\ 1 & 0\end{array}\right).
\end{array}
$$ 
We first describe our scheme for $n=2$.
If the user wants to know $x_1$, he builds the following 3-qubit state
$$
\frac{1}{\sqrt{2}}\left(\ket{0}\ket{B_{00}}+\ket{1}\ket{B_{01}}\right),
$$
and if he wants to know $x_2$ he builds
$$
\frac{1}{\sqrt{2}}\left(\ket{0}\ket{B_{00}}+\ket{1}\ket{B_{10}}\right).
$$
He sends the second qubit to server~1 and the third to server~2,
keeping the first qubit to himself. It is easy to see that each server
always gets a completely mixed qubit, so the servers learn nothing about $i$.
Both servers will now apply $\sigma_{x_1x_2}$ to the qubit they receive.
That is, they will apply a phase flip if $x_1=1$ and a bit flip if $x_2=1$.
The following properties are easily verified:
$$
\begin{array}{rcr}
(\sigma_x\otimes\sigma_x)\ket{B_{00}} & = & \ket{B_{00}}\\
(\sigma_x\otimes\sigma_x)\ket{B_{01}} & = & (-1)^{x_1}\ket{B_{01}}\\
(\sigma_x\otimes\sigma_x)\ket{B_{10}} & = & (-1)^{x_2}\ket{B_{10}}
\end{array}
$$
The servers then send their qubit back to the user.
By the above properties, if the user wanted to know $x_1$, then he now has
$$
\frac{1}{\sqrt{2}}\left(\ket{0}\ket{B_{00}}+(-1)^{x_1}\ket{1}\ket{B_{01}}\right),
$$
and if he wanted $x_2$ he has
$$
\frac{1}{\sqrt{2}}\left(\ket{0}\ket{B_{00}}+(-1)^{x_2}\ket{1}\ket{B_{10}}\right).
$$
{}From this the user can extract the bit $x_i$ of his choice 
(with probability~1)---and nothing else.
Thus we have an honest-user 2-server QSPIR for $n=2$ with $4$ qubits of communication.

To generalize to arbitrary $n=2m$, the user can employ a larger state
that involves $m$ Bell states to extract $x_i$. 
Namely, if $i=2j-1$ ($1\leq j\leq m$) then he uses
$$
\frac{1}{\sqrt{2}}\left(\ket{0}\ket{B_{00}}^{\otimes m}+
\ket{1}\ket{B_{00}}^{\otimes j-1}\ket{B_{01}}\ket{B_{00}}^{\otimes m-j}\right),
$$
and if $i=2j$ then he uses
$$
\frac{1}{\sqrt{2}}\left(\ket{0}\ket{B_{00}}^{\otimes m}+
\ket{1}\ket{B_{00}}^{\otimes j-1}\ket{B_{10}}\ket{B_{00}}^{\otimes m-j}\right).
$$
The user sends the left qubit of each of the Bells states to server~1,
the right qubit of each Bell state to server~2, and keeps 
the first qubit to himself.
The servers then apply $\sigma_{x_{2j-1}x_{2j}}$ to the $j$th qubit 
they receive (for all $1\leq j\leq m$) and send back the result.
Using the same properties as before, it can easily be verified 
that we just get the appropriate phase-factor $(-1)^{x_i}$ in 
the $\ket{1}$-part of the user's total state and nothing else.
Thus we have a scheme that works for all $n$ and that simultaneously
hides $i$ from the servers and $x-x_i$ from an honest user.
In total, the scheme uses $2n$ qubits of communication: 
$m=n/2$ to each server, and $m=n/2$ back. 

\section{Dishonest-user quantum SPIR schemes}

The assumption that the user is honest (i.e., follows the protocol)
is somewhat painful, since the servers cannot rely on this.
In particular, a dishonest quantum user can extract about $\log n$
bits of information about $x$ of any honest-user QSPIR where
the user's final state is pure, as follows. Consider such a pure 
QSPIR scheme, with as many servers and communication as you like. 
From the user's high level perspective, this can be viewed 
as a unitary that maps
$$
\ket{i}\ket{0}\rightarrow \ket{i}\ket{x_i}\ket{\phi_{i,x_i}}.
$$
Because of data privacy, the state $\ket{\phi_{i,x_i}}$ only depends 
on $i$ and $x_i$. Therefore by one application of the QSPIR and some 
unitary post-processing, the user can erase $\ket{\phi_{i,x_i}}$,
mapping
$$
\ket{i}\ket{0}\rightarrow \ket{i}\ket{x_i},
$$
for any $i$ or superposition of $i$s of his choice.
That is, one run of the QSPIR can be used to make one {\em query\/} to $x$.
Van Dam~\cite{dam:oracle} has shown how one quantum 
query to $x$ can be used to obtain $\Omega(\log n)$ bits of information 
about $x$ (in the information-theoretic sense that is, 
not necessarily $\log n$ specific database-bits $x_j$). 
Accordingly, any pure QSPIR that is secure against an honest user will leak
at least $\Omega(\log n)$ bits of information about $x$ to a cheating user.
This includes our schemes from the previous section. 
Even worse, the servers cannot even {\em detect\/} whether the user cheats,
because they will have the same state in the honest scheme 
as well as in the cheating scheme.

How to protect against dishonest quantum users?
In fact we can just use a classical SPIR that is secure against
dishonest users (of course, this will be a shared-randomness scheme again).  
If we require the servers to measure what they
receive in the computational basis, then a dishonest quantum user
cannot extract more information than a classical dishonest 
user---that is, nothing except one $x_i$.

\section{Conclusion}

We have shown that the best known PIR schemes can be turned into
quantum PIR schemes that are symmetrically private with respect to
a honest user, i.e., except for the bit $x_i$ that he asks for,
the honest user receives no information whatsoever about the 
database $x$. Rather interestingly, the best known {\em quantum\/}
PIR schemes use polynomially less communication than the best
known classical schemes (Table~\ref{tablepir}), 
but our PIR-to-QSPIR reduction does not seem to work 
starting from a quantum PIR system. We leave it as an open
question whether the communication complexity of QSPIR schemes
can be significantly reduced, either based on the 
QPIR schemes of~\cite{kerenidis&wolf:qldc} or via some other method.


\end{document}